\documentclass{PoS}

\usepackage{wrapfig}

\newcommand{\be}{\begin{equation}}
\newcommand{\ee}{\end{equation}}
\newcommand{\U}{\textnormal{U}}
\newcommand{\Og}{\textnormal{O}}
\newcommand{\SU}{\textnormal{SU}}
\newcommand{\scan}[1]{{\bf #1}}
\newcommand{\refc}[1]{(\ref{#1})}
\newcommand{\Op}{\mathcal{O}}

\title{Testing the strength of the $\U_A(1)$ anomaly at the chiral phase transition in two-flavour QCD}

\ShortTitle{Strength of the $\U_A(1)$ anomaly at the $N_f=2$ chiral phase transition}

\author{\speaker{Bastian B. Brandt}, Owe Philipsen\\%\thanks{A footnote may follow.}\\
        Institute f\"ur Theoretische Physik, Goethe Universit\"at, Max-von-Laue-Strasse 1, 60438 Frankfurt am Main, Germany \\
        E-mail: \email{brandt@th.physik.uni-frankfurt.de}\\}

\author{Marco C\`e\\
        Helmholtz-Institut Mainz, Johannes Gutenberg-Universit\"at Mainz, Germany\\}
        
\author{Anthony Francis\\
        Theoretical Physics Department, CERN, CH-1211 Geneva 23, Switzerland\\}

\author{Tim Harris\\
        Dip. di Fisica G. Occhialini, Universit\`a di Milano-Bicocca, and INFN, Sezione di
        Milano-Bicocca, Piazza della Scienza 3, I-20126 Milano, Italy\\}

\author{Harvey B. Meyer, Hartmut Wittig\\
        PRISMA Cluster of Excellence and Institut f\"ur Kernphysik and Helmholtz-Institut Mainz,
        Johannes Gutenberg-Universit\"at Mainz, Germany}

\abstract{We study the thermal transition of QCD with two degenerate light flavours by lattice
simulations using $\mathcal{O}(a)$-improved Wilson quarks. Particular emphasis lies on the pattern of
chiral symmetry restoration, which we probe via the static screening correlators. On $32^3$ volumes we
observe that the screening masses in transverse iso-vector vector and axial-vector channels become degenerate
at the transition temperature. The splitting between the screening masses in iso-vector scalar and
pseudoscalar channels is strongly reduced compared to the splitting at zero temperature and is actually
consistent with zero within uncertainties. In this proceedings article we extend our studies to matrix
elements and iso-singlet correlation functions. Furthermore, we present results on larger volumes, including
first results at the physical pion mass.}

\FullConference{The 9th International workshop on Chiral Dynamics\\
		17-21 September 2018\\
		Durham, NC, USA}

\begin{document}

\section{Introduction}

One of the peculiar features of Quantum Chromodynamics (QCD) is the spontaneous and explicit breaking
of chiral symmetry. In the case with two massless quark flavours the theory is invariant under global
transformations with elements of
\be
\SU_L(2) \times \SU_R(2) \times \U_A(1) \times \U_V(1) \,.
\ee
The $\U_A(1)$ symmetry, however, is anomalously broken on the quantum level due to the Adler-Bell-Jackiw,
or chiral, anomaly. In particular, the conservation equation for the axial vector current reads
\be
\partial_\mu A^j_\mu(x) = - \delta_{j0} \frac{N_f\,g_0^2}{32\pi^2}
\varepsilon^{\alpha\beta\mu\nu} \textnormal{Tr}\big[F_{\alpha\beta}(x)  F_{\mu\nu}(x)\big]
= - \delta_{j0} N_f Q(x) \,,
\ee
where $Q(x)$ denotes the operator associated with the topological charge density and
\be
A^j_\mu(x) = \bar{\psi}(x) \gamma_\mu \gamma_5 (\tau^j/2) \psi(x) \,,
\ee
is the axial current, including the Pauli matrices $\tau^j$ for $j=1,2,3$ and $\tau^0=\mathbf{1}$.
Due to this anomalous breaking,
the spontaneous breaking of chiral symmetry leads to three Goldstone bosons only, neutral and charged
pions, while the $\eta'$ meson retains a finite mass, even for vanishing light quark
masses~\cite{Witten:1979vv,Veneziano:1979ec}.

\begin{figure}[t]
 \centering
 \vspace*{-3mm}
\includegraphics[width=.8\textwidth]{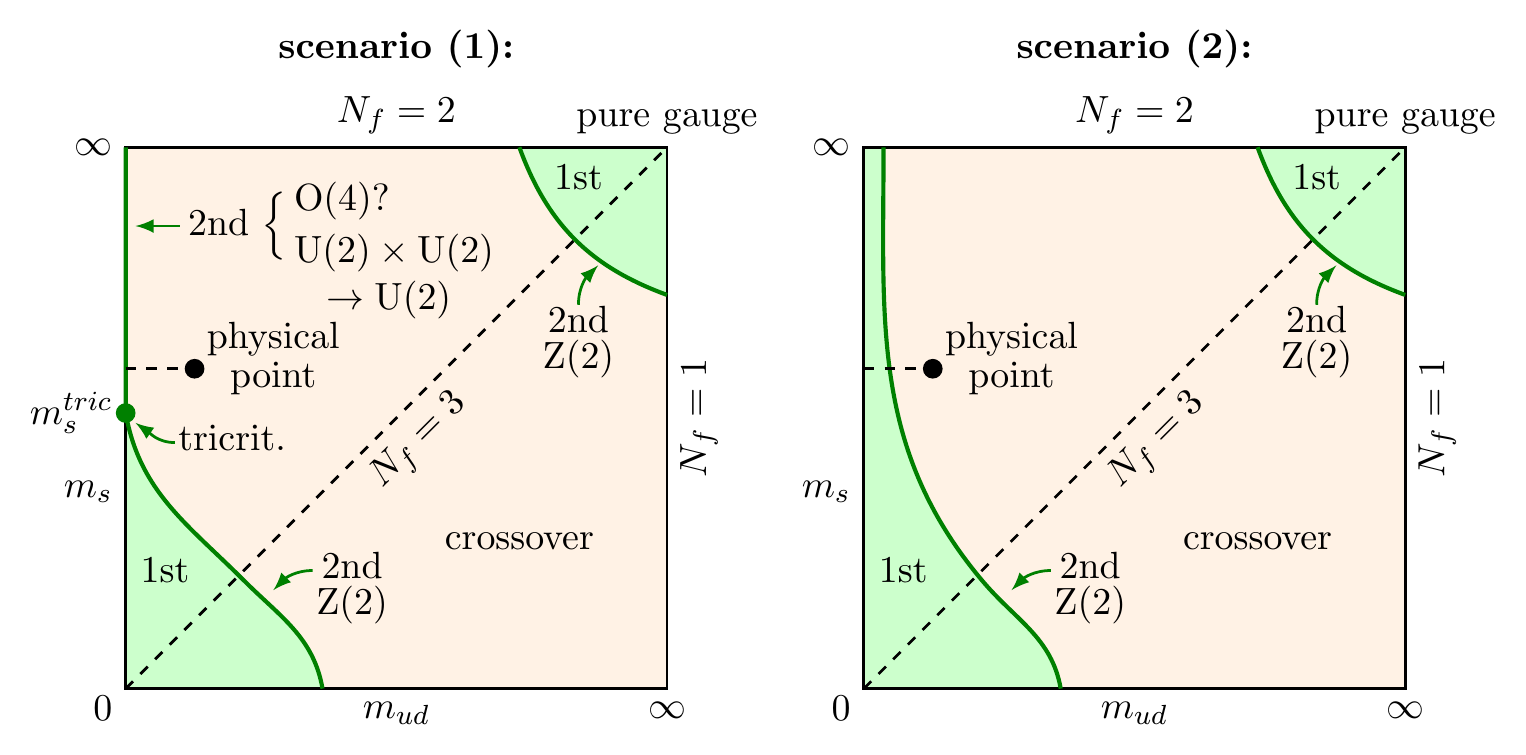}
 \vspace*{-3mm}
 \caption{The two possible scenarios for the phase structure of QCD at zero chemical potential.}
 \label{fig:columbia}
\end{figure}

The fate of the anomalous breaking of the $\U_A(1)$ symmetry at finite temperature plays a key role for
the properties of the QCD phase diagram. In particular, the phase transition in the chiral limit of the
light ($u$ and $d$) quarks is sensitive to the possible restoration of the $\U_A(1)$ symmetry at the
critical temperature, which could change the order and/or the universality class of the transition
(see Refs.~\cite{Pisarski:1983ms,Butti:2003nu,Pelissetto:2013hqa}). The two possible scenarios for the
QCD phase diagram in dependence of the masses of the three lightest quarks are shown in
Fig.~\ref{fig:columbia}. In scenario (1), the 2nd order chiral critical line reaches the $m_{ud}=0$ axis
in a tricritical point at $m_s^{tric}$, rendering the chiral transition 2nd order from this point on.
The universality class in this scenario depends on the strength of the breaking of $\U_A(1)$ at the chiral
transition. If the effect of the breaking is negligible, i.e. the symmetry effectively restored, the
transition will be in the $\U(2) \times \U(2) \to \U(2)$~\cite{Butti:2003nu,Pelissetto:2013hqa}
universality class (alternatively a $\SU(2)\times\SU(2)\times Z_4\to\SU(2)$ universality class has also been
proposed~\cite{Aok:2016gxw}) rather than in the standard $\Og(4)$ universality
class~\cite{Pisarski:1983ms} for a substantial breaking of $\U_A(1)$. It is
also possible that the restoration of the $\U_A(1)$ symmetry is sufficient to keep the transition
first order for all values of the strange quark mass~\cite{Pisarski:1983ms}. This is scenario (2) in
Fig.~\ref{fig:columbia}. The question which of the two scenarios is realised is the only remaining
completely open qualitative question of the phase diagram at vanishing chemical potential. Among the
main problems to answer this questions are the inability to simulate directly in the chiral limit
and the similarity of the different types of scaling behaviour. Investigating the fate of the
$\U_A(1)$ symmetry at the chiral phase transition offers a viable alternative to the above
methods (see our earlier paper~\cite{Brandt:2016daq} for a more detailed discussion and references).

\begin{figure}[b]
 \centering
 \vspace*{-3mm}
\includegraphics[width=.58\textwidth]{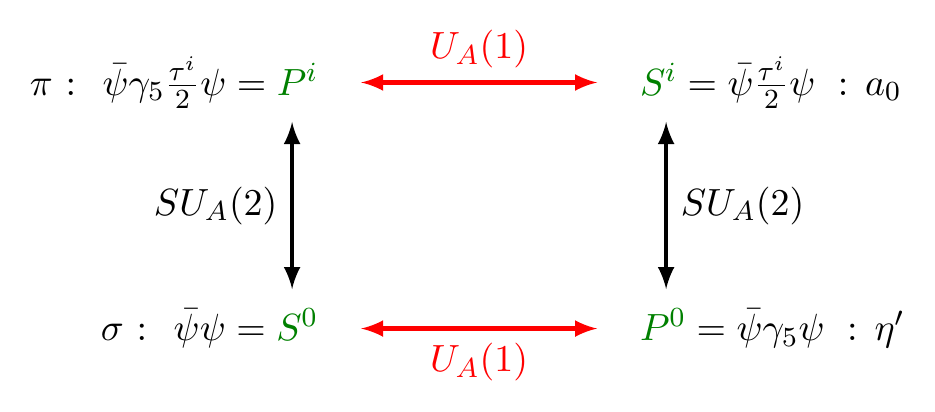}
 \caption{Transformation relations between iso-vector and iso-singlet operators in $P$ and $S$ channels.}
 \label{fig:UA1-pattern}
\end{figure}

The pattern of chiral symmetry restoration can be investigated using correlation functions of operators
connected by the individual symmetries (see also~\cite{GomezNicola:2017bhm}).
One particular example are iso-vector correlation functions in vector
$V^j_\mu(x)=\bar{\psi}(x) \gamma_\mu \gamma_5 (\tau^j/2) \psi(x)$ and axial vector $A^j_\mu(x)$ channels,
which are related by the $\SU_A(2)$ rotation. Consequently, the restoration of chiral symmetry
implies the degeneracy of the associated correlation functions. Of particular relevance for the $\U_A(1)$
symmetry are correlation functions in scalar and pseudoscalar channels. Including the iso-singlet
operators (opening up new channels for the investigation of the effective symmetry restoration,
see Ref.~\cite{Nicola:2018vug}, for instance), they are related by $\SU_A(2)$ and $\U_A(1)$
transformations as shown in Fig.~\ref{fig:UA1-pattern}.
The iso-vector operators $P^i$ and $S^i$ are comparably easy to compute on the lattice, due to the
absence of disconnected diagrams, so that they have become the standard channels to look at to test for
$\U_A(1)$ symmetry restoration. Since we are considering an effective restoration of the symmetry,
we expect the renormalised correlation functions to become degenerate.

A number of studies have looked at the effective restoration of the $\U_A(1)$ symmetry in lattice QCD,
mostly focussing on the low mode spectrum of the Dirac operator or chiral
susceptibilities~\cite{Bazavov:2012qja,Cossu:2013uua,Chiu:2013wwa,Dick:2015twa,Tomiya:2016jwr}.
In contrast, we pursue a complementary
approach, using the correlation functions, in particular, the screening masses. Screening masses
probe the long distance properties of the correlation functions and are free of contact terms, which
contaminate chiral susceptibilities, for instance. Apart from screening masses, the correlation
functions include additional information in terms of matrix elements. The details of our strategy 
are explained in~\cite{Brandt:2016daq}, where the iso-vector screening masses obtained from
$16\times32^3$ lattices have been published. Here we extend this study to larger volumes, the matrix
elements of the correlation functions and present first results for iso-singlet screening masses,
which provide additional information about the symmetry restoration pattern. Earlier accounts of our
study have been reported in~\cite{Brandt:2010uw,Brandt:2010bn,Brandt:2012sk,Brandt:2013mba}.

\section{Simulation Setup}

We perform simulations with two flavours of $O(a)$ improved Wilson
fermions~\cite{Sheikholeslami:1985ij}, with the non-perturbatively estimated clover coefficient 
from Ref.~\cite{Jansen:1998mx}, and Wilson's gauge action. The simulations employ deflation
accelerated versions of the Schwarz~\cite{Luscher:2005rx,Luscher:2007es} and
(twisted) mass~\cite{Hasenbusch:2001ne,Marinkovic:2010eg,Luscher:2012av} preconditioned
algorithms. We vary the temperature by changing the lattice spacing via the
lattice coupling $\beta$, keeping the temporal extent fixed at $N_t=16$.
For more details concerning the simulation algorithms, lines of
constant physics and scale setting see~\cite{Brandt:2016daq}. In the approach to the
chiral limit we use several quark masses and volumes to control finite size effects.
A list of temperature scans with the results for the critical temperatures is
given in Tab.~\ref{tab:scans}

\begin{table}[t]
\begin{center}
\begin{tabular}{cc|ccccc}
\hline
\hline
scan & Volume & $m_{ud}$~[MeV] & $m_\pi$~[MeV] & $T_c$~[MeV] & $\beta_c$ & cfg \\
\hline
\scan{B1}$_\kappa$ & $32^3$ & $\sim41$ & $\sim485$ & 232(19) & 5.465 &
$\sim1000$ \\
\hline
\scan{C1} & $32^3$ & $\sim17.5$ & 300 &  211( 6) & 5.405 & $\sim400$ \\
\scan{C2} & $48^3$ &  &  &  &  & $\sim1000$ \\
\hline
\scan{D1} & $32^3$ & $\sim8.7$ & 220 & 190(12) & 5.340 & $\sim750$ \\
\scan{D2} & $48^3$ &  &  &  &  & $\sim800$ \\
\hline
\scan{E2} & $48^3$ & $\sim3.6$ & 135 & $\approx183$ & 5.317 & $\sim400$ \\
\hline
\hline
\end{tabular}
\end{center}
\caption{$\beta$-scans at $N_t=16$. Listed are the lattice volume in lattice
units, the quark mass $m_{ud}$, the zero-temperature pion mass $m_\pi$
(estimated via NNLO $\chi$PT, see~\cite{Brandt:2016daq}), the critical
temperature $T_c$, the critical lattice coupling $\beta_c$ and the approximate
number of independent configurations
per ensemble `cfg' (estimated via the integrated autocorrelation time of the
plaquette at $T_c$). Scan \scan{B1}$_\kappa$ has been done with a
constant hopping parameter $\kappa$ rather than at constant $m_{ud}$.
The quoted quark and pion masses correspond to the ones at $T_c$.
For scan \scan{E2} the critical temperature has been estimated from the $\Og(4)$
scaling fit to the $32^3$ lattices~\cite{Brandt:2016daq}.}
\label{tab:scans}
\end{table}

Our main observables are correlation functions in spatial direction, so called
screening correlators~\cite{DeTar:1987ar}. For a particular mesonic
operator $O$ the screening correlation function is given by
\be
\label{eq:meson-corrfunc}
C_{O}(x_\mu) = \int d^3x_\perp \left< O(x_\mu,\vec{x}_\perp)O^\dagger(0) \right> \,.
\ee
Here $x_\mu$ (we take $x_\mu=z$) is the coordinate of the direction
in which the correlation function is evaluated and $\vec{x}_\perp$ is the coordinate
vector in the orthogonal directions. The equality of the renormalised correlation
functions of channels related by a particular symmetry signals its effective restoration.
Previously we have focussed on correlation functions
with iso-vector operators, i.e.\ operators including a Pauli matrix $\tau^i$, for
which only quark connected correlation functions contribute. Here we will also
present first results for quark disconnected correlation functions, enabling us to
compute correlation functions in iso-singlet channels. The details will be discussed
in Sec.~\ref{sec:disc}.

On a periodic lattice of extent $L_z$, the leading order of the spectral representation,
including only the groundstate contribution, of a correlation function $C_{O}(z)$ is given by
\be
\label{eq:corr-func-spect}
C_{O}(z) = \frac{\big| Z_O \big|^2}{M_O}
\Big( e^{-M_O z} + e^{-M_O (L_z-z)} \Big) \,.
\ee
The exponential decay of $C_{O}(x_\mu)$ with $x_\mu$ defines the
`screening mass' $M_O$ in this channel and the proportionality constant contains
the matrix element $Z_O$. Consequently, the equality of the correlation functions
not only implies the equivalence of the screening masses, but also of the
renormalised matrix elements $\mathcal{Z}_O Z_O$, where $\mathcal{Z}_O$ are the
multiplicative renormalisation factors.

\section{Anomalous Breaking of $\U_A(1)$ from Iso-Vector Correlation Functions}

We start with the discussion of the results for correlation functions in the
iso-vector channels. In particular, we look at the correlation functions
from Eq.~\refc{eq:meson-corrfunc} in pseudoscalar $P^i$, scalar $S^i$, vector
$V^i$ and axial vector $A^i$ channels. In this section we
conveniently drop the superscript $i$ for brevity. Iso-vector correlation functions
include a connected part only, which we evaluate using point sources. We typically
use 48 point sources per configuration with random starting positions. We
first focus on the screening mass differences,
\be
\Delta M_{O_1 O_2} = M_{O_1} - M_{O_2} \,,
\ee
which are direct measures for the effective restoration of the symmetries.
These differences are extracted from plateaus in the effective masses of the
ratios of the correlation functions in the individual channels, taking into account
the leading order contributions from excited states.

\begin{figure}[t]
 \centering
 \vspace*{-3mm}
\includegraphics[width=.8\textwidth]{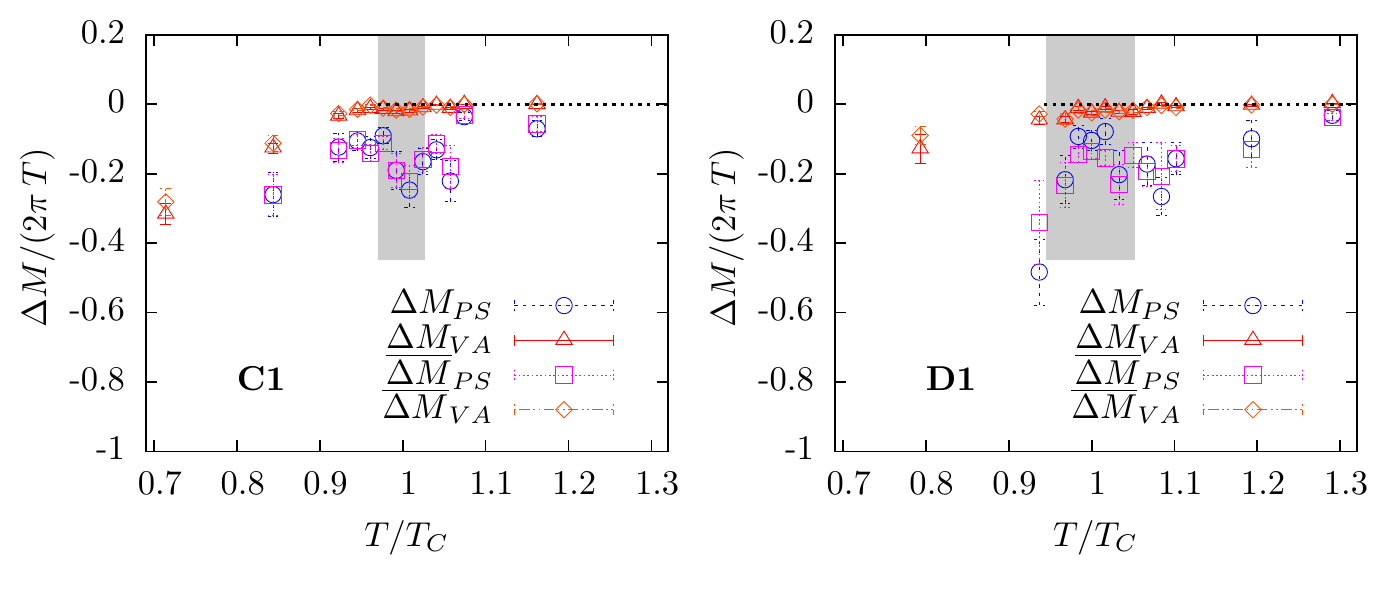}
 \caption{Results for the differences $\Delta M_{PS}$ and $\Delta M_{VA}$,
 for scans \scan{C1} (left) and \scan{D1} (right). The differences are
 normalised to $2\pi T$. The grey area marks the transition region.
 The results indicated by $\overline{\Delta M}$
 are an alternative estimate for the screening mass difference
 (see~\cite{Brandt:2016daq}).}
 \label{fig:scm-dif-paper}
\end{figure}

\begin{figure}[t]
 \centering
\includegraphics[width=.8\textwidth]{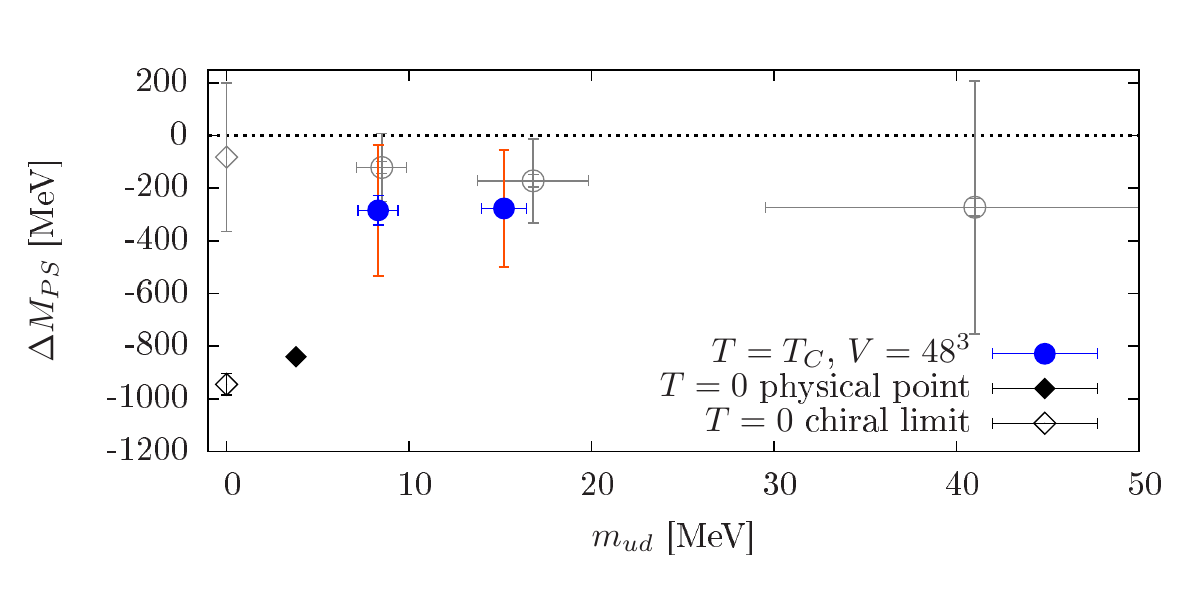}
 \caption{Results for $\Delta M_{PS}$ from the temperature scans
 with a volume of $48^3$. The grey points are the results from
 the $32^3$ volumes for comparison and the black
 points are the reference values at $T=0$ (see text).}
 \label{fig:ps-diff}
\end{figure}

The results for the screening mass differences on the $32^3$ volumes, scans
\scan{C1} and \scan{D1}, as obtained in~\cite{Brandt:2016daq}, are shown in
Fig.~\ref{fig:scm-dif-paper}. We observe an approximate degeneracy of $M_V$
and $M_A$, indicating the effective restoration of $\SU_A(2)$ at $T_c$.
The difference $\Delta M_{PS}$ is non-vanishing at $T_c$, meaning $\U_A(1)$
is still broken for these quark masses. This finding is in qualitative
and quantitative agreement with findings from staggered~\cite{Cheng:2010fe},
domain wall~\cite{Bazavov:2012qja,Bhattacharya:2014ara,Cossu:2014aua} and
overlap~\cite{Cossu:2013uua,Tomiya:2014mma} fermion formulations. In the
approach to the chiral limit the difference decreases. To obtain an estimate
in the chiral limit we perform a linear chiral extrapolation of $\Delta M_{PS}$
(averaging over the transition region and using the spread of results as a
systematic uncertainty; see Ref.~\cite{Brandt:2016daq}). The results from
the averaging procedure for scans \scan{B1}$_\kappa$, \scan{C1} and
\scan{D1} versus the quark mass are shown as the grey points in
Fig.~\ref{fig:ps-diff} together with their chiral extrapolation, the grey
point at $m_{ud}=0$. To enable an assessment whether the
breaking is weak or strong in the chiral limit, Fig.~\ref{fig:ps-diff}
also includes a phenomenologically estimate for the mass difference in the
chiral limit and at the physical point in full QCD at
$T=0$~\cite{Brandt:2016daq}. A comparison between the chiral extrapolation
and the phenomenological estimate shows that the $\U_A(1)$ breaking
screening-mass difference is comparably small at $T_c$, indicating a weak
breaking or even a restoration of the $\U_A(1)$ symmetry at $m_{ud}=0$.

\begin{figure}[t]
 \centering
 \vspace*{-3mm}
\includegraphics[width=.8\textwidth]{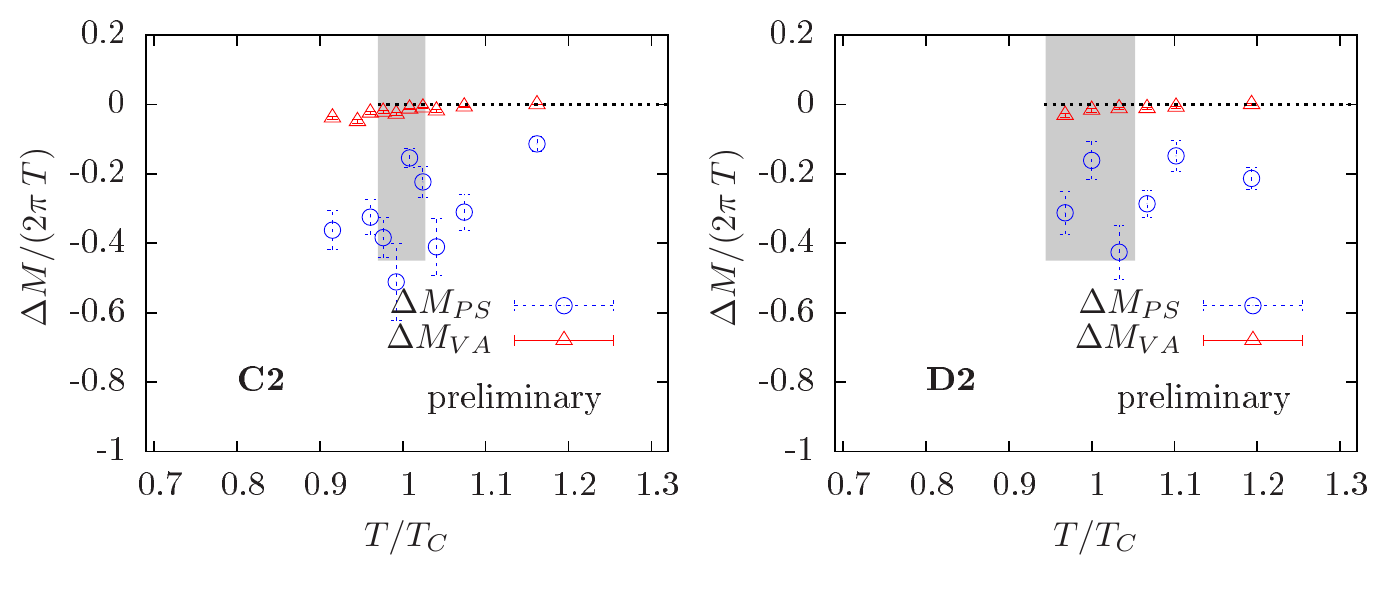}
 \caption{Results for the differences $\Delta M_{PS}$ and $\Delta M_{VA}$,
 for scans \scan{C2} (left) and \scan{D2} (right).}
 \label{fig:scm-diff}
\end{figure}

For the $(L/a)^3=32^3$ volumes and smaller quark masses, the value for $m_\pi L$,
with $m_\pi$ the $T=0$ pion mass, becomes smaller than 3. To be able to extend
the study to smaller quark masses and to test for finite size effects, we have
thus repeated the computations on $48^3$ lattices. The results for the screening
mass differences on these new scans \scan{C2} and \scan{D2} are shown in
Fig.~\ref{fig:scm-diff}. While the results for $\Delta M_{VA}$ look very similar to
the ones from the $32^3$ volumes, $\Delta M_{PS}$ tends to become larger
with increasing volume. The result for $\Delta M_{PS}$ at $T_c$, once more
averaged over the transition region, are shown in Fig.~\ref{fig:ps-diff}. One
can see the tendency towards larger screening mass differences with increasing
volume. This tendency seems to remain for the chiral limit. To perform a
reliable chiral extrapolation, however, we need to extend the simulations to
smaller quark masses and increase the statistics for scan \scan{D2}.

\begin{wrapfigure}{r}{6.5cm}
 \centering
 \vspace*{-3mm}
 \includegraphics[width=6.5cm]{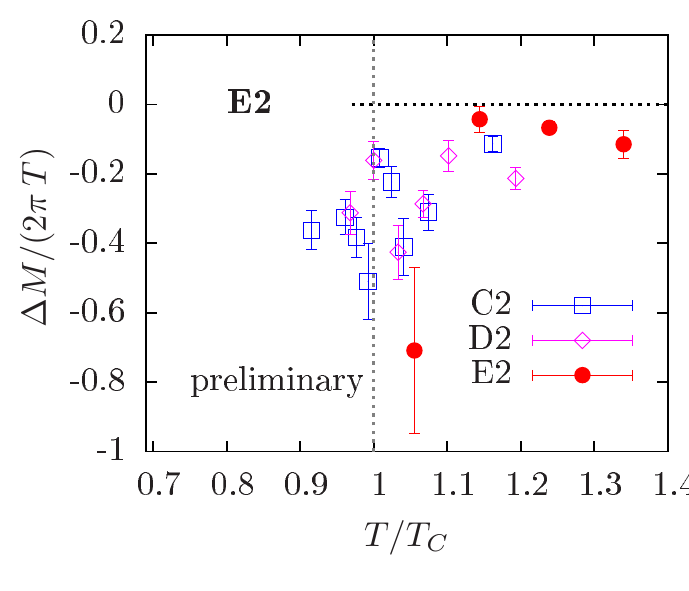}
 \caption{Results for the differences $\Delta M_{PS}$, for scans \scan{C2},
 \scan{D2} and \scan{E2}.}
 \label{fig:scm-diff-phys}
\end{wrapfigure}

To extend our study to smaller quark masses we have started a
temperature scan at the physical pion mass, labelled \scan{E2} in
Tab.~\ref{tab:scans}. So far only results at $T>T_c$ ($T_c$ estimated using
$\Og(4)$ scaling~\cite{Brandt:2016daq}) are available, for which we have performed
measurements with 16 point sources per configuration. The results are shown
in Fig.~\ref{fig:scm-diff-phys}. While the results for $T/T_c>1.1$ indicate a smaller
 value for $\Delta M_{PS}$ the one at $T/T_c\approx1.05$ is further
away from zero as for scans \scan{C2} and \scan{D2} at similar values for $T/T_c$.
The latter, however, lacks statistics and thus can still change considerably.
We are currently increasing precision and are extending the runs to $T_c$ and below.

\begin{figure}[b]
 \centering
 \vspace*{-3mm}
\includegraphics[width=.8\textwidth]{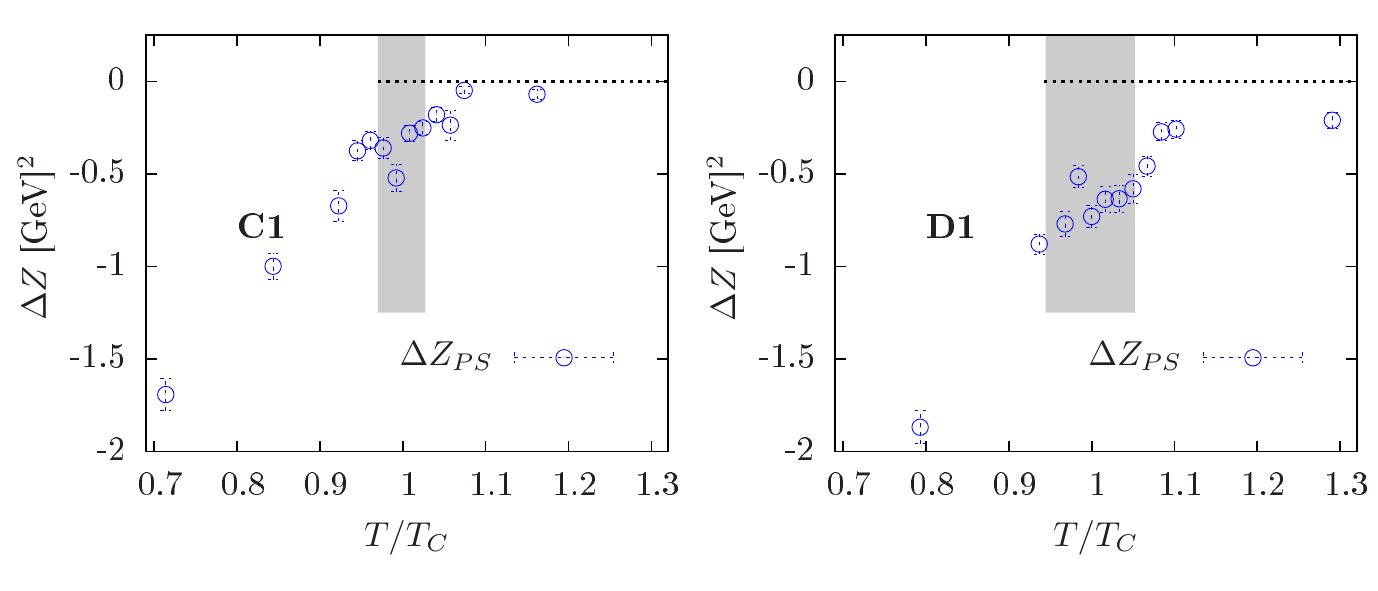}
 \caption{Results for the differences $\Delta Z_{PS}$, for scans \scan{C2}
 and \scan{D2}.}
 \label{fig:mat-diff}
\end{figure}

An alternative observable extracted from correlation functions are
the matrix elements $Z_O$ from Eq.~\refc{eq:corr-func-spect}. In contrast to the
screening masses, however, these observables demand multiplicative
renormalisation. The analogue to $\Delta M_{PS}$ for the matrix elements is the
renormalised difference
\be
\Delta Z_{PS} = \mathcal{Z}_P |Z_P| - \mathcal{Z}_S |Z_S| \,.
\ee
For the determination of $\mathcal{Z}_P$, we have
interpolated the results from~\cite{Fritzsch:2012wq} as discussed
in~\cite{Brandt:2016daq}. The determination of $\mathcal{Z}_S$ is a bit more
involved and we refer to the renormalisation of the chiral condensate
in~\cite{Brandt:2016daq} for the details. The renormalised difference is plotted
for scans \scan{C1} and \scan{D1} in Fig.~\ref{fig:mat-diff}. As the screening mass
difference, $\Delta Z_{PS}$ remains non-zero at $T_c$. In contrast to
$\Delta M_{PS}$, however, $\Delta Z_{PS}$ shows a tendency to increase when
the quark mass is lowered.

\section{Iso-Singlet Screening Correlators}
\label{sec:disc}

\begin{wrapfigure}{r}{6.5cm}
 \centering
 \vspace*{-5mm}
 \includegraphics[]{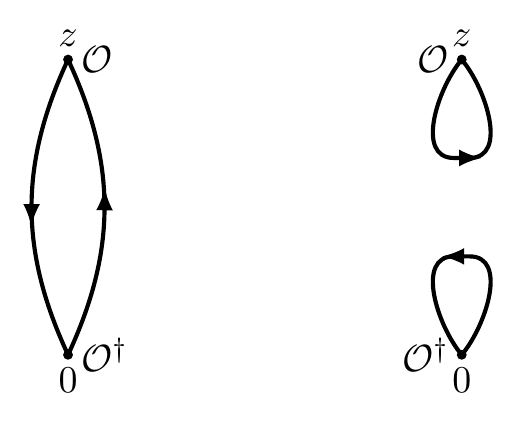}
 \caption{Graphical representation of quark connected (left) and disconnected
 (right) diagrams in lattice QCD.}
 \label{fig:conn-disc-diags}
\end{wrapfigure}

Iso-singlet correlation functions open up new channels to investigate the chiral
symmetry restoration pattern. In two-flavour QCD with degenerate quark masses, the
difference between iso-vector and iso-singlet correlation function is the presence
of quark disconnected diagrams for the latter. In this section we will distinguish
explicitly between iso-vector $O^i$ and iso-singlet $O^0$ operators/correlation
functions. Introducing quark connected $C^{\rm conn}_{\Op}(z)$ and quark
disconnected $C^{\rm disc}_{\Op}(z)$ correlation functions, defined by traces over
quark propagators as depicted in Fig.~\ref{fig:conn-disc-diags}, iso-vector and
iso-singlet correlation functions of an operator $O$ are given by,
\be
C_{O^i}(z) = - \frac{1}{2} C^{\rm conn}_{\Op}(z) \quad \textnormal{and} \quad
C_{O^0}(z) = - \frac{1}{2} C^{\rm conn}_{\Op}(z) + C^{\rm disc}_{\Op}(z) \,.
\ee
Here $\Op$ refers to the operator in Dirac and colour space only, excluding
the flavour matrices from the operator $O$, and the correlation functions
include only a single fermion propagator (since $u$ and $d$ propagators are
indistinguishable in QCD when quark masses are degenerate). Note, that
the iso-singlet correlation function might include a constant
contribution, which is a finite size effect resulting from imperfect sampling
of topological sectors~\cite{Aoki:2007ka}. The constant piece is absent
in the shifted correlator $\tilde{C}(z)=C(z)-C(z+1)$~\cite{Ottnad:2017bjt}, which
we use to fit the correlation function in the $P^0$ channel.

We will focus on correlation functions in $P$ and $S$ channels at $T_c$.
The transformation relations between iso-vector and iso-singlet channels are
shown in Fig.~\ref{fig:UA1-pattern}. We see that the inclusion of the disconnected
diagrams enables to test the effective restoration of both symmetries using
$P$ and $S$ correlation functions. In particular, in the case that $\SU_A(2)$
is restored, for which we have seen indications above,
we expect a degeneracy of $P^0$ and $S^i$, as well as $P^i$ and $S^0$
correlation functions. If, the $\U_A(1)$ symmetry remains
broken, correlation functions in $P^i$ and $S^i$ channels and $P^0$ and
$S^0$ channels remain non-degenerate.

\begin{figure}[b]
 \centering
 \vspace*{-3mm}
\includegraphics[width=.4\textwidth]{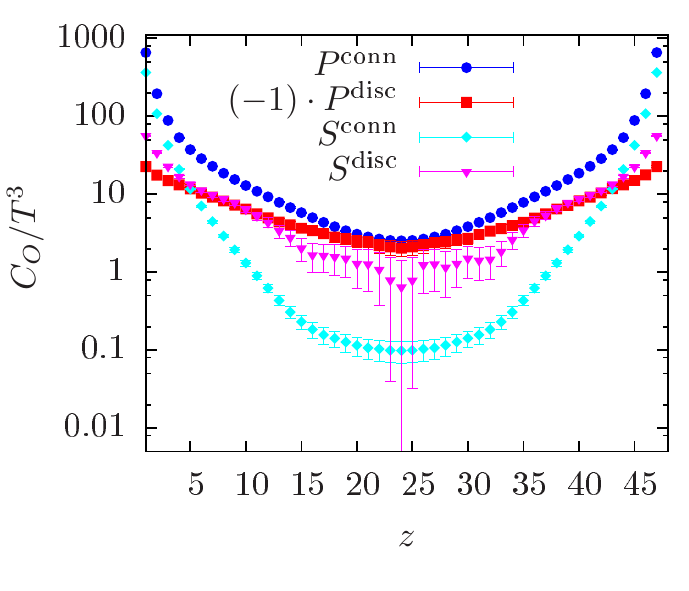} \hfil
\includegraphics[width=.4\textwidth]{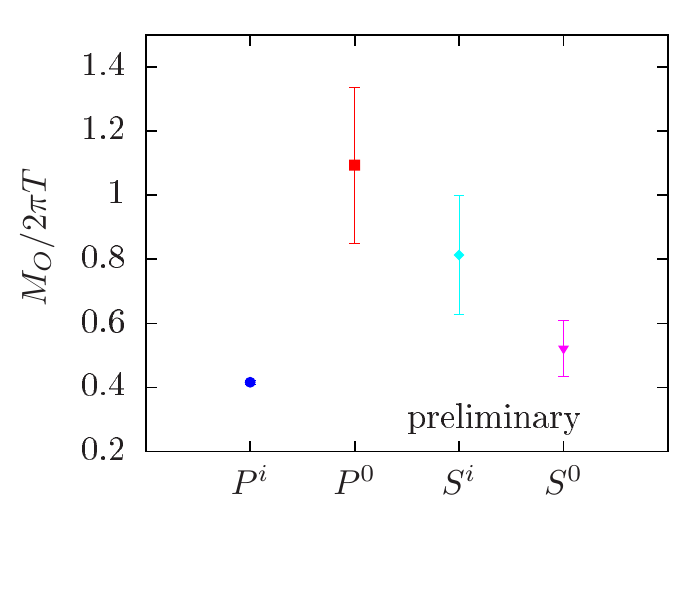}
 \caption{Screening correlation functions (left) and masses (right) for the
 different channels at the critical temperature of scan \scan{C2}.}
 \label{fig:disc-corrs}
\end{figure}

We will first compare the connected and disconnected parts of the correlation
functions, shown in the left panel of Fig.~\ref{fig:disc-corrs}.
The disconnected correlation functions have been computed with 32
Hadamard probing vectors for hierarchical probing~\cite{Stathopoulos:2013aci}.
While the magnitude of connected and disconnected correlation functions in
the $P$ channel is similar, the disconnected correlator is negative, leading
to a large cancellation. At $T=0$ this cancellation results in the exponential
decay of the iso-singlet correlator with the $\eta'$ mass. In the $S$ channel,
both correlation functions are positive, but the disconnected correlation
function has a slower exponential decay and thus governs the iso-singlet
correlator. To investigate the pattern of chiral symmetry restoration,
we have extracted the iso-singlet screening masses from the correlators,
which we compare to the iso-vector screening masses in the right panel of
Fig.~\ref{fig:disc-corrs}. Both, the screening masses in $P^i$ and $S^0$ channels,
as well as in $P^0$ and $S^i$ channels agree within (their large) uncertainties,
confirming the effective restoration of $\SU_A(2)$. At the same time, also the
screening masses in $P^0$ and $S^0$ channels are non-degenerate, indicating
a residual breaking of $\U_A(1)$, as seen in the previous section.

Note, that the results are preliminary in the sense that currently the statistics
is not sufficient to extract the iso-singlet screening masses reliably.
In particular, the screening masses in
Fig.~\ref{fig:disc-corrs} (except for $M_{P^i}$) have been extracted without
taking excited states into account. This is problematic since the signals
become lost in noise already at comparably small values of $z$. The obvious
next task is to increase statistics and to confirm the results presented in
this section.

\section{Conclusions}

In this proceedings article we have updated our initial study~\cite{Brandt:2016daq}
to larger volumes, showed first results for a scan at physical light quark masses,
extended our set of observables and presented first results for the extension to
iso-singlet correlation functions. Larger volumes show the tendency to strengthen
the effect of the anomalous breaking of the $\U_A(1)$ symmetry. However, a chiral
extrapolation of the screening mass difference at $T_c$ is currently not possible
for the larger volumes, lacking a third quark mass. We are currently extending
our simulations to the physical quark mass, for which we have shown first results.
Additional observables are renormalised matrix elements of the screening correlators.
Evaluated on the $32^3$ volumes, they tend to show an increase in the strength
of the breaking of $\U_A(1)$ for smaller quark masses. A first look at screening
masses in iso-singlet channels at $T_c$ for scan \scan{C2} shows that the breaking of
$\U_A(1)$ is also present in $P^0$ and $S^0$ channels, while invariance under
$\SU_A(2)$ transformations appears to be restored. To be able to reliably extract
information from the iso-singlet correlators, however, a substantial increase
in precision is mandatory.

\noindent{\bf Acknowledgements:} \\
B.B. would like to thank J. Ruiz de Elvira and Y. Aoki for stimulating
discussions and the conveners of the GB Dynamics session for the
invitation to present the results. We acknowledge computer time for
the generation of the gauge configurations on the JUROPA, JURECA and JUQUEEN
computers of the Gauss Centre for Supercomputing at FZ J\"ulich,
allocated through the John von Neumann Institute for Computing (NIC) within project
HMZ21. Parts of the simulations have been done on
``Wilson'' at the Institute for Nuclear Physics, University of Mainz,
``Clover'' at the Helmholtz-Institut Mainz and the FUCHS cluster
at the Centre for Scientific Computing, University of Frankfurt.
This work has been supported by the Cluster of Excellence PRISMA+ (EXC 2118/1),
funded by DFG within the German Excellence Strategy (Project ID 39083149),
the {\em Center for Computational Sciences in Mainz} as part of the Rhineland-Palatine
Research Initiative and by DFG grant ME 3622/2-2 {\em QCD at finite temperature with
Wilson fermions on fine lattices}. B.B. has also received funding by the DFG via
SFB/TRR 55 and the Emmy Noether Programme EN 1064/2-1.

\end{document}